\documentclass[%
 reprint,
 superscriptaddress,
 amsmath,amssymb,
 aps,
]{revtex4-2}

\usepackage{graphicx}% Include figure files
\usepackage{dcolumn}% Align table columns on decimal point
\usepackage{bm}% bold math
\usepackage{xcolor}
\usepackage{algorithm} 
\usepackage{algpseudocode} 

\begin{document}

%\preprint{APS/123-QED}

\title{Temporal scaling theory for bursty time series with clusters of arbitrarily many events}

\author{Hang-Hyun Jo}
\email{h2jo@catholic.ac.kr}
\affiliation{Department of Physics, The Catholic University of Korea, Bucheon, Republic of Korea}

\author{Tibebe Birhanu}
\affiliation{Department of Physics, The Catholic University of Korea, Bucheon, Republic of Korea}

\author{Naoki Masuda}
\affiliation{
Department of Mathematics, State University of New York at Buffalo, USA
}
\affiliation{
Institute for Artificial Intelligence and Data Science, State University of New York at Buffalo, USA
}
\affiliation{
Center for Computational Social Science, Kobe University, Kobe, Japan}

\date{\today}% It is always \today, today,

\begin{abstract}
Long-term temporal correlations in time series in a form of an event sequence have been characterized using an autocorrelation function (ACF) that often shows a power-law decaying behavior. Such scaling behavior has been mainly accounted for by the heavy-tailed distribution of interevent times (IETs), i.e., the time interval between two consecutive events. Yet little is known about how correlations between consecutive IETs systematically affect the decaying behavior of the ACF. Empirical distributions of the burst size, which is the number of events in a cluster of events occurring in a short time window, often show heavy tails, implying that arbitrarily many consecutive IETs may be correlated with each other. In the present study, we propose a model for generating a time series with arbitrary functional forms of IET and burst size distributions. Then, we analytically derive the ACF for the model time series. In particular, by assuming that the IET and burst size are power-law distributed, we derive scaling relations between power-law exponents of the ACF decay, IET distribution, and burst size distribution. These analytical results are confirmed by numerical simulations. Our approach helps to rigorously and analytically understand the effects of correlations between arbitrarily many consecutive IETs on the decaying behavior of the ACF.
\end{abstract}

\maketitle

\section{Introduction}\label{sec:intro}

Complex systems often show complex dynamical behaviors such as long-term temporal correlations, also known as $1/f$ noise~\cite{Hooge1981Experimental, Weissman1988Noise, Bak1987Selforganized, Bak1996How, Allegrini2009Spontaneous}. Characterizing those temporal correlations is of utmost importance to understand mechanisms behind such observations. There are a number of characterization and measurement methods in the literature; e.g., one can refer to Refs.~\cite{Kantz2004Nonlinear, Kantelhardt2012Fractal, Karsai2018Bursty, Masuda2021Guide} and references therein. One of the most commonly used measurements is an autocorrelation function (ACF)~\cite{Fano1950Shorttime, Kantelhardt2001Detecting}. Precisely, the ACF for a time series $x(t)$ is defined with a time lag $t_{\rm d}$ as
\begin{align}
    A(t_{\rm d})\equiv \frac{ \langle x(t)x(t+t_{\rm d})\rangle_t- \langle x(t)\rangle^2_t}{ \langle x(t)^2\rangle_t- \langle x(t)\rangle^2_t},
    \label{eq:acf_define}
\end{align}
where $\langle\cdot\rangle_t$ is the time average over the entire period of the time series. The ACF has been extensively used for detecting temporal correlations in various natural and social phenomena~\cite{Brunetti1984Analysis, Koscielny-Bunde1998Longrange, Min2005Observation, Karsai2018Bursty}. For the time series with long-term correlations, the ACF typically decays in a power-law form with a decay exponent $\gamma$ such that
\begin{align}
    A(t_{\rm d})\propto t_{\rm d}^{-\gamma}.
    \label{eq:gamma}
\end{align}
This power-law behavior is closely related to the $1/f$ noise via Wiener-Khinchin theorem~\cite{Leibovich2016Aging} as well as to the Hurst exponent $H$~\cite{Hurst1956Problem} and its generalizations~\cite{Peng1994Mosaic, Barunik2010Hurst, Rybski2009Scaling, Rybski2012Communication, Tang2015Complexity}.

A type of time series that has attracted attention is given in a form of a sequence of event timings or an event sequence, which can be regarded as realizations of point processes in time~\cite{Daley2003Introduction}. Temporal correlations in such event sequences have been characterized by the time interval between two consecutive events, namely, an interevent time (IET). In many empirical data sets, IET distributions, denoted by $P(\tau)$, have heavy tails~\cite{Bak2002Unified, Corral2004LongTerm, Barabasi2005Origin, deArcangelis2006Universality, Vazquez2006Modeling, Bedard2006Does, Bogachev2007Effect, Malmgren2008Poissonian, Malmgren2009Universality, Kemuriyama2010Powerlaw, Wu2010Evidence, Tsubo2012Powerlaw, Kivela2015Estimating, Gandica2017Stationarity}; in particular, $P(\tau)$ often shows a power-law tail as
\begin{align}
    P(\tau)\propto \tau^{-\alpha}
    \label{eq:alpha}
\end{align}
with a power-law exponent $\alpha$~\cite{Karsai2018Bursty}. Lowen and Teich derived the analytical solution of the power spectral density for a renewal process governed by a power-law IET distribution~\cite{Lowen1993Fractal, Lowen2005Fractalbased}, concluding that the decay exponent $\gamma$ in Eq.~\eqref{eq:gamma} is solely determined by the IET exponent $\alpha$ in Eq.~\eqref{eq:alpha} as
\begin{align}
    \gamma=\begin{cases}
    2-\alpha & \textrm{for}\ 1<\alpha<2,\\
    \alpha-2 & \textrm{for}\ 2<\alpha<3.
    \end{cases}
    \label{eq:PRE1993}
\end{align}
It is not surprising because the heavy-tailed IET distribution is the only source of temporal correlations in the time series for renewal processes, where there are no correlations between IETs. The same scaling relation in Eq.~\eqref{eq:PRE1993} was derived in other model studies~\cite{Abe2009Violation, Vajna2013Modelling, Lee2018Hierarchical}.

In general, correlations between IETs, in addition to the IET distribution, should also be relevant to the understanding of asymptotic decay of the ACF. To detect correlations between consecutive IETs, a notion of bursty trains was introduced~\cite{Karsai2012Universal}; for a given time window $\Delta t$, consecutive events are clustered into a bursty train when any two consecutive events in the train are separated by the IET smaller than or equal to $\Delta t$, while the first (last) event in the train is separated from the last (first) event in the previous (next) train by the IET larger than $\Delta t$. The number of events in each bursty train is called a burst size, and it is denoted by $b$. By analyzing various data, Karsai et al.~reported that the burst size distribution shows power-law tails as
\begin{align}
    Q_{\Delta t}(b)\propto b^{-\beta}
    \label{eq:beta}
\end{align}
with a power-law exponent $\beta$ for a wide range of $\Delta t$~\cite{Karsai2012Universal}. Similar observations have been made in other data~\cite{Karsai2012Correlated, Yasseri2012Dynamics, Jiang2013Calling, Kikas2013Bursty, Wang2015Temporal, Jo2020Bursttree}. These findings immediately raise an important question: how does the ACF decay power-law exponent $\gamma$ depend on the IET power-law exponent $\alpha$ as well as on the burst-size power-law exponent $\beta$?

It is not straightforward to devise a model or process that can answer this question, because the heavy tail of the burst size distribution typically implies that arbitrarily many consecutive IETs may be correlated with each other. It is worth noting that correlations between two consecutive IETs have been quantified in terms of the memory coefficient~\cite{Goh2008Burstiness}, local variation~\cite{Shinomoto2003Differences}, and mutual information~\cite{Baek2008Testing}; they were implemented using, e.g., a copula method~\cite{Jo2019Copulabased, Jo2019Analytically} and a correlated Laplace Gillespie algorithm in the context of many-body systems~\cite{Masuda2018Gillespie, Masuda2022Gillespie}. Correlations between an arbitrary number of consecutive IETs have been modeled by means of, e.g., the two-state Markov chain~\cite{Karsai2012Universal}, self-exciting point processes~\cite{Jo2015Correlated}, the IET permutation method~\cite{Jo2017Modeling}, and a model inspired by the burst-tree decomposition method~\cite{Jo2020Bursttree}. Although scaling behaviors of the ACF were studied in some of mentioned works~\cite{Karsai2012Universal, Jo2015Correlated, Jo2017Modeling, Jo2019Analytically}, the scaling relation $\gamma(\alpha,\beta)$ has not been clearly understood due to the lack of analytical solutions of the ACF.

\begin{figure}[!t]
\includegraphics[width=0.8\columnwidth]{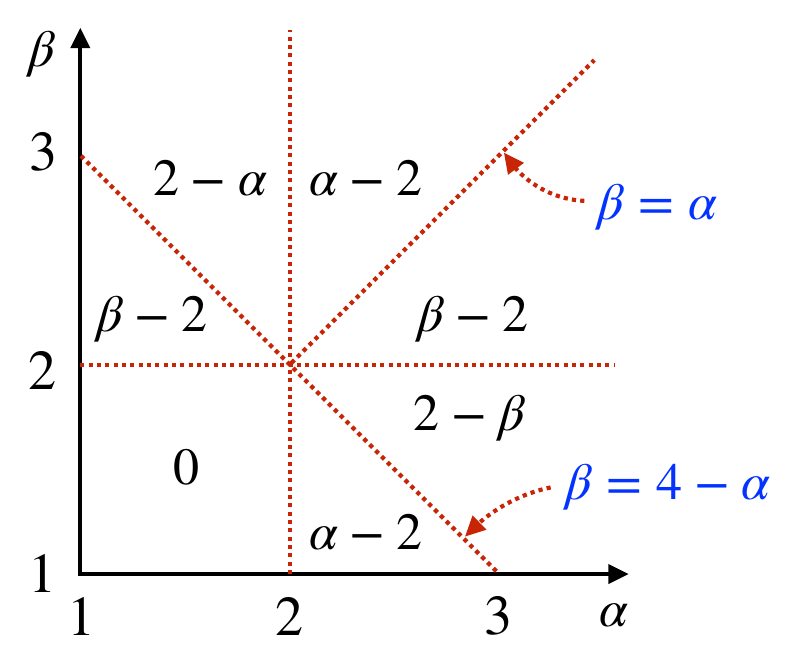}
\caption{Visualization of the main result, i.e., $\gamma(\alpha,\beta)$, given by Eq.~\eqref{eq:gamma_summary}. Here $\gamma$ is a power-law exponent for the decaying behavior of the autocorrelation function, $\alpha$ is a power-law exponent of the interevent time distribution, and $\beta$ is a power-law exponent of the burst size distribution.}
\label{fig:main_result}
\end{figure}

In this work, we devise a model for generating a time series with arbitrary functional forms of the IET distribution and burst size distribution. Assuming power-law tails for IET and burst size distributions, our model generates correlations between an arbitrary number of consecutive IETs. By theoretically analyzing the model, we derive asymptotically exact solutions of the ACF from the model time series, enabling us to find the scaling relation $\gamma(\alpha,\beta)$ as follows:
\begin{align}
    \gamma=\begin{cases}
    0 & \textrm{for}\ 1<\alpha,\beta\leq 2\ \textrm{or}\ \alpha=2\ \textrm{or}\ \beta=2,\\
    \alpha-2 & \textrm{for}\ \beta>\alpha>2\ \textrm{or}\ \alpha>2,\ \beta<4-\alpha,\\
    2-\alpha & \textrm{for}\ 1<\alpha<2,\ \beta>4-\alpha,\\
    \beta-2 & \textrm{for}\ \alpha>\beta>2\ \textrm{or}\ \beta>2,\ \beta<4-\alpha,\\
    2-\beta & \textrm{for}\ 1<\beta<2,\ \beta> 4-\alpha.
    \end{cases}
    \label{eq:gamma_summary}
\end{align}
We depict the result in Fig.~\ref{fig:main_result}. Note that, for the case of $\beta\gg 1$, the scaling relation in Eq.~\eqref{eq:PRE1993} is partly recovered.

The paper is organized as follows. In Sec.~\ref{sec:model}, we introduce the model with arbitrary functional forms of IET and burst size distributions. In Sec.~\ref{sec:analysis}, we provide an analytical framework for the derivation of the ACF for the model time series. In Sec.~\ref{sec:powerlaw}, by assuming power-law distributions of IETs and burst sizes, we derive analytical solutions of the ACF, hence the decay exponent $\gamma$ as a function of $\alpha$ and $\beta$. We also compare the obtained analytical results with numerical simulations. Finally, we conclude our paper in Sec.~\ref{sec:conclusion}.

\section{Model}\label{sec:model}

Let us introduce the following model for generating event sequences $\{x(1), \ldots, x(T)\}$ in discrete time, where $T$ is the number of discrete times, for given interevent time (IET) distribution and burst size distribution. By definition, $x(t)=1$ if an event occurs at time $t\in \{1, \ldots, T\}$, and $x(t)=0$ otherwise. By construction, the minimum IET is $\tau_{\rm min}=1$. Furthermore, we only consider bursts with $\Delta t=1$ throughout this work. In other words, we regard that events occurring in consecutive times form a burst, including the case in which the burst contains only one event.

\begin{figure}[!t]
\includegraphics[width=\columnwidth]{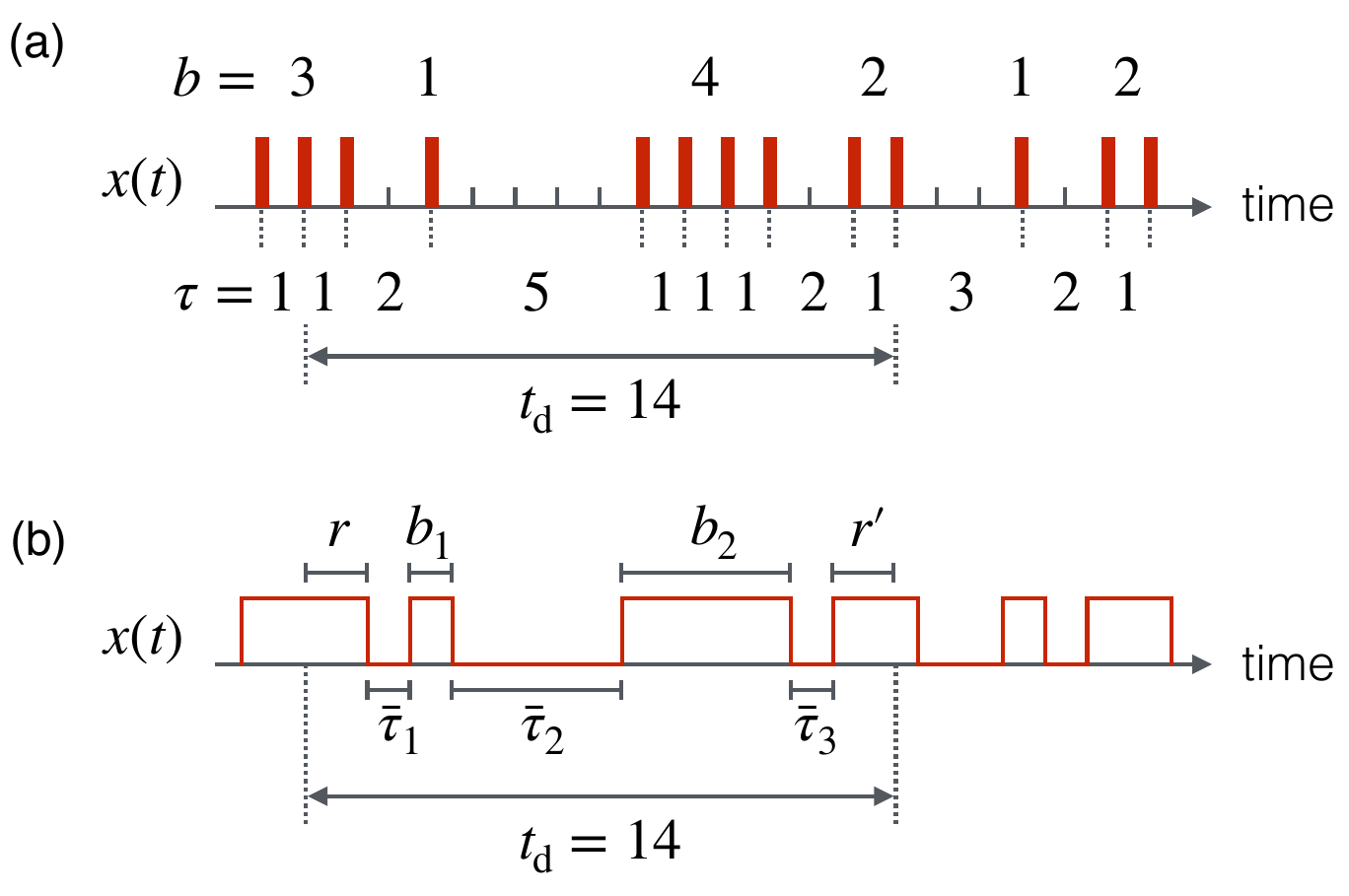}
\caption{(a) Schematic diagram for an event sequence $x(t)$ in discrete time $t$, showing how interevent times $\tau$ and burst sizes $b$ are combined in the event sequence, with an example of non-zero $x(t)x(t+t_{\rm d})$ for the time lag $t_{\rm d}=14$. (b) The event sequence $x(t)$ can be seen as the alternation of periods of $x(t)=1$ and those of $x(t)=0$. A typical composition of the time lag $t_{\rm d}$ for non-zero $x(t)x(t+t_{\rm d})$ is presented. See the main text for the definitions of symbols.}
\label{fig:scheme}
\end{figure}

The event sequence $x(t)$ is generated using an IET distribution for $\tau \ge 2$, denoted by $\psi(\tau)$, and a burst size distribution $Q(b)$. Note that $\psi(\tau)$ is normalized. To generate an event sequence, we first randomly draw a burst size $b_1$ from $Q(b)$ to set $x(t)=1$ for $t=1,\ldots,b_1$. Then, we randomly draw an IET $\tau_1$ from $\psi(\tau)$ to set $x(t)=0$ for $t=b_1+1,\ldots,b_1+\tau_1-1$. Note that $\tau_1 \ge 2$. We draw another burst size $b_2$ from $Q(b)$ and another IET $\tau_2$ from $\psi(\tau)$, respectively, to set $x(t)=1$ for $t=b_1+\tau_1,\ldots,b_1+\tau_1+b_2-1$ and $x(t)=0$ for $t=b_1+\tau_1+b_2,\ldots,b_1+\tau_1+b_2+\tau_2-1$. We repeat this procedure until $t$ reaches $T$. See Fig.~\ref{fig:scheme}(a) for an example. 

\subsection{Remarks}

We have two remarks that are related to each other. Firstly, in our model, all IETs of $\tau=1$ are generated by bursts with $b\geq 2$ events; a burst of size $b$ generates exactly $b-1$ IETs of $\tau=1$. This is why the fraction of IETs of $\tau=1$ among all IETs is determined by $Q(b)$. To be precise, if there are $m$ bursts in the generated event sequence, there must be the asymptotically same number of IETs of $\tau\geq 2$ in the limit of $T,m\to \infty$. Then the number of IETs of $\tau=1$ is $m(\langle b\rangle -1)$, where $\langle b\rangle$ is the average burst size, i.e.,
\begin{align}
    \langle b\rangle \equiv \sum_{b=1}^\infty bQ(b).
\end{align}
The fraction of IETs of $\tau\geq 2$, denoted by $u$, is obtained as
\begin{align}
    u=\frac{m}{m(\langle b\rangle -1)+m}=\frac{1}{\langle b\rangle}.
\label{eq:u=1/b}
\end{align}
Using this $u$, one can write the IET distribution for the entire range of IETs as follows:
\begin{align}
    P(\tau)=(1-u)\delta(\tau,1)+u[1-\delta(\tau,1)]\psi(\tau),
    \label{eq:Ptau_u}
\end{align}
where $\delta(\cdot,\cdot)$ is a Kronecker delta. 

The second remark is for the general case including our model; in general, $P(\tau)$ and $Q_{\Delta t}(b)$ are not independent of each other. Following Refs.~\cite{Jo2017Modeling, Jo2018Limits, Jo2023Bursty}, let us consider a sequence of $n$ events, from which one obtains $n-1$ IETs. For a given time window $\Delta t$, $n$ events are clustered into, say, $m$ bursty trains, implying that the average burst size is given by
\begin{align}
    \langle b\rangle_{\Delta t} \equiv \sum_{b=1}^\infty bQ_{\Delta t}(b)=\frac{n}{m}.
    \label{app_eq:b_avg}
\end{align}
Here $m$ is closely related to the number of IETs larger than $\Delta t$ as follows:
\begin{align}
    m=(n-1)F(\Delta t)+1,
    \label{app_eq:m_n}
\end{align}
where $F(\Delta t)\equiv\int_{\Delta t}^\infty \text{d}\tau P(\tau)$ is a cumulative probability distribution function of $P(\tau)$. By combining Eqs.~\eqref{app_eq:b_avg} and~\eqref{app_eq:m_n}, one obtains
\begin{align}
    \frac{1}{\langle b\rangle_{\Delta t}}=\left(1-\frac{1}{n}\right)F(\Delta t)+\frac{1}{n}.
\end{align}
In the limit of $n\to\infty$, we get
\begin{align}
    \langle b\rangle_{\Delta t} F(\Delta t)\simeq 1.
    \label{app_eq:b_avg_Ftau}
\end{align}
If we assume pure power-law forms of $P(\tau)$ in Eq.~\eqref{eq:alpha} and of $Q_{\Delta t}(b)$ in Eq.~\eqref{eq:beta}, we can derive the scaling relation between $\alpha$ and $\beta$, possibly undermining our main question about $\gamma(\alpha,\beta)$. However, most empirical distributions of IETs and burst sizes are not pure power-laws~\cite{Karsai2018Bursty}, enabling us to treat $\alpha$ and $\beta$ as independent parameters to some extent. Later we will assume that the IET distribution has a power-law tail for $\tau\geq 2$, while the burst size distribution is a pure power-law.

\section{Analytical framework\label{sec:analysis}}

We analyze the autocorrelation function (ACF) given by Eq.~\eqref{eq:acf_define} for the time series $\{x(1), \ldots, x(T)\}$ generated by the model described in Sec.~\ref{sec:model}. Because $A(t_{\rm d}=0)=1$ by definition, we consider the positive time lag ($t_{\rm d}>0$) unless we state otherwise. For an event sequence composed of $n$ events, one gets the event rate as 
\begin{equation}
\lambda\equiv \langle x(t)\rangle_t = \frac{n}{T},
\end{equation}
enabling to write $\langle x(t)^2\rangle_t=\langle x(t)\rangle_t=\lambda$ because $x(t) \in \{ 0, 1 \}$. The term $\langle x(t)x(t+t_{\rm d})\rangle_t$ in Eq.~\eqref{eq:acf_define} is written as follows:
\begin{align}
    \langle x(t)x(t+t_{\rm d})\rangle_t&=\Pr[x(t)=1\wedge x(t+t_{\rm d})=1]\nonumber\\
    &=\Pr[x(t)=1]\cdot \Pr[x(t+t_{\rm d})=1|x(t)=1]\nonumber\\
    &=\lambda Z(t_{\rm d}),
\end{align}
where $Z(t_{\rm d})$ is the probability that $x(t+t_{\rm d})=1$ conditioned on $x(t)=1$. Note that $Z(0) = 1$. Then, the ACF reads
\begin{align}
    A(t_{\rm d})=\frac{Z(t_{\rm d})-\lambda}{1-\lambda}.
    \label{eq:acf_simple}
\end{align}

Let us consider the case in which $x(t)x(t+t_{\rm d})$ is non-zero, i.e., $x(t)=1$ and $x(t+t_{\rm d}) = 1$. As depicted in Fig.~\ref{fig:scheme}(a), the time series $x(t)$ in a period of $[t,t+t_{\rm d}]$ is typically composed of several alternating bursts and IETs larger than one. Here the consecutive IETs of $\tau=1$ forms a burst and the sum of such IETs of $\tau=1$ is equal to the burst size minus one. Therefore, the time lag $t_{\rm d}$ is written as a sum of burst sizes (each minus one) of bursts appeared in the period of $[t,t+t_{\rm d}]$ and IETs between those bursts. We note that the first burst contains an event at time $t$ and that the last burst contains an event at time $t+t_{\rm d}$.

We denote by $r$ the number of events from time $t$ to the end of the burst containing the event at time $t$. By definition, we exclude the event at time $t$ when counting $r$, hence $r\geq 0$. For example, if a burst is composed of $5$ events occurring at times $1, \ldots, 5$ and we consider $t=2$, then one obtains $r=3$. If we select $t$ such that $x(t)=1$ uniformly at random, $t$ is contained in a burst of size $b$ with a probability proportional to $Q(b)$. Therefore, $r$ is a discrete-time variant of the waiting time or the residual time derived from the IET~\cite{Karsai2018Bursty}, and the probability distribution of $r$ is given by
\begin{align}
    R(r)= \frac{1}{\langle b\rangle} \sum_{b=r+1}^\infty Q(b).
    \label{eq:R_define}
\end{align}
Note that $\sum_{r=0}^{\infty} R(r) = 1$. See Appendix~\ref{append:CR} for the derivation of the factor $1/\langle b\rangle$.

Next we denote by $r'$ the number of events that are in the burst containing the event at time $t+t_{\rm d}$ and occur before or at $t+t_{\rm d}$ [Fig.~\ref{fig:scheme}(b)]. It should be noted that $r'\geq 1$. The definition of $r'$ implies that the first event of the burst containing the event at $t + t_{\rm d}$ occurred at time $t+t_{\rm d} - r' + 1$. We denote by $q(r')$ the probability that $x(t+t_{\rm d}) = 1$ conditioned that the first event of the burst containing the event at time $t + t_{\rm d}$ is located at $t+t_{\rm d} - r' + 1$. For this case to occur, the size of the burst starting at time $t+t_{\rm d} - r' + 1$ has to be larger than or equal to $r'$. Therefore, one obtains
\begin{align}
    q(r')= \sum_{b=r'}^\infty Q(b).
    \label{eq:Rf}
\end{align}

To derive $Z(t_{\rm d})$ in Eq.~\eqref{eq:acf_simple}, we denote by $p_k(t_{\rm d})$ the probability that an event occurs at time $t+t_{\rm d}$ and there are exactly $k$ alterations between the burst and the IET larger than one, conditioned that an event occurs at time $t$. By $k$ alterations, we mean that the event at time $t+t_{\rm d}$ belongs to the $k$th burst after the burst to which the event at time $t$ belongs to. Note that there are then $k$ IETs larger than one between the burst at time $t$ and that at time $t+t_{\rm d}$. Then $Z(t_{\rm d})$ can be written in terms of $p_k(t_{\rm d})$ as
\begin{align}
    Z(t_{\rm d})=\sum_{k=0}^\infty p_k(t_{\rm d}).
    \label{eq:Z_define}
\end{align}

For the case with $k=0$, we obtain using Eq.~\eqref{eq:R_define}
\begin{align}
    p_0(t_{\rm d})= \sum_{r = t_{\rm d}}^\infty R(r) = \frac{1}{\langle b\rangle} \sum_{b=t_{\rm d}+1}^\infty Q(b)(b-t_{\rm d}).
    \label{eq:p_0}
\end{align}
It should be noted that, when counting $t_{\rm d}$ for $k=0$, we exclude the event at time $t$ and include the event at time $t+t_{\rm d}$ for consistency. For example, consider a burst of $5$ events at times $1, \ldots,5$. If we are considering the two events at times 2 and 4, we set $t=2$ and $t_{\rm d} = 2$.

If there are $k$ $(\ge 1)$ IETs larger than one intersecting $[t,t+t_{\rm d}]$, one can write [Fig.~\ref{fig:scheme}(b)]: 
\begin{align}
t_{\rm d} &= r+ \sum_{i=1}^k \tau_i+\sum_{i=1}^{k-1}(b_i-1)+r'-1\notag\\
&= r+ \sum_{i=1}^k\bar\tau_i+\sum_{i=1}^{k-1}b_i+r',
\label{eq:td_define}
\end{align}
where we have defined a reduced IET by
\begin{align}
    \bar\tau_i \equiv \tau_i - 1
    \label{eq:bartau_define}
\end{align}
for convenience. Note that $\bar{\tau}_i \ge 1$ for all $i$ because $\tau_i\geq 2$. We assume that all variables on the right-hand side of Eq.~\eqref{eq:td_define} are statistically independent of each other. Then, we can write $p_k(t_{\rm d})$ for $k\geq 1$ as
\begin{align}
    p_k(t_{\rm d})=
    &\sum_{r=0}^\infty \left(\prod_{i=1}^k \sum_{\bar\tau_i=1}^\infty\right) \left(\prod_{i=1}^{k-1} \sum_{b_i=1}^\infty\right) \sum_{r'=1}^\infty\nonumber\\
    &R(r) \left[\prod_{i=1}^k \psi(\bar\tau_i)\right] \left[\prod_{i=1}^{k-1}Q(b_i)\right] q(r') \nonumber\\
    &\times \delta\left(t_{\rm d},  
    r + \sum_{i=1}^k\bar\tau_i + \sum_{i=1}^{k-1}b_i + r'\right).
    \label{eq:p_k}
\end{align}
It is also remarkable that using the reduced IET, the event rate is obtained as follows:
\begin{align}
    \lambda= \langle x(t)\rangle_t = \frac{\langle b\rangle}{\langle\bar\tau\rangle+\langle b\rangle},
    \label{eq:lambda}
\end{align}
where
\begin{align}
    \langle\bar\tau\rangle\equiv \sum_{\bar\tau=1}^\infty \bar\tau\psi(\bar\tau).
\end{align}
Namely, $\lambda$ is the ratio of the average length of the period of $x(t)=1$ to the sum of those of $x(t)=1$ and of $x(t)=0$ [Fig.~\ref{fig:scheme}(b)].

For analytical tractability, we assume that all variables on the right-hand side of Eq.~\eqref{eq:td_define} are real numbers. Therefore, $\psi(\bar\tau)$, $Q(b)$, $R(r)$, $q(r')$, and $p_0(t_{\rm d})$ are also considered for their respective continuous variables. The continuous versions of Eqs.~\eqref{eq:R_define} and \eqref{eq:Rf} are given by
\begin{align}
    \label{eq:R_cont_define}
    &R(r)=\frac{1}{\langle b\rangle}\int_{r}^\infty \text{d}b Q(b)
\end{align}
and
\begin{align}
    \label{eq:Rf_cont_define}
    &q(r')= \int_{r'}^\infty \text{d}b Q(b).
\end{align}
respectively. Furthermore, the continuous-time versions of Eqs.~\eqref{eq:p_0}~and~\eqref{eq:p_k} respectively read 
\begin{align}
    p_0(t_{\rm d})=\frac{1}{\langle b\rangle} \int_{t_{\rm d}}^\infty \text{d}b Q(b)(b-t_{\rm d})
    \label{eq:p_0_cont}
\end{align}
and
\begin{align}
    p_k(t_{\rm d})=
    &\int_0^\infty \text{d}r \left(\prod_{i=1}^k \int_0^\infty \text{d}\bar\tau_i\right) \left(\prod_{i=1}^{k-1} \int_0^\infty \text{d}b_i\right) \int_0^\infty \text{d}r' \nonumber\\
    &R(r) \left[\prod_{i=1}^k \psi(\bar\tau_i)\right] \left[\prod_{i=1}^{k-1}Q(b_i)\right] q(r') \nonumber\\
    &\times \delta\left(t_{\rm d}- 
    r- \sum_{i=1}^k\bar\tau_i-\sum_{i=1}^{k-1}b_i-r'\right),
    \label{eq:p_k_cont}
\end{align}
where $\delta(\cdot)$ is the Dirac delta function. 

We take the Laplace transform of Eq.~\eqref{eq:p_0_cont} to obtain 
\begin{align}
    \tilde p_0(s)=\frac{1}{s}-\frac{1-\tilde Q(s)}{\langle b\rangle s^2},
    \label{eq:p0s}
\end{align}
where $\tilde Q(s)$ denotes the Laplace transform of $Q(b)$. The Laplace transform of Eq.~\eqref{eq:p_k_cont} reads for $k\geq 1$
\begin{align}
    \tilde p_k(s)=\tilde R(s)\tilde q(s) \tilde \psi(s)^k \tilde Q(s)^{k-1},
\end{align}
where 
\begin{align}
    &\tilde R(s)=\frac{1-\tilde Q(s)}{\langle b\rangle s},
    \label{eq:Rs}\\
    &\tilde q(s) = \frac{1-\tilde Q(s)}{s}
    \label{eq:qs},
\end{align}
and $\tilde \psi(s)$ denotes the Laplace transform of $\psi(\bar\tau)$. Then the Laplace transform of $Z(t_{\rm d})$ in Eq.~\eqref{eq:Z_define} is obtained as
\begin{align}
    \tilde Z(s)=\sum_{k=0}^\infty \tilde p_k(s)
    = \tilde p_0(s)+\frac{\tilde R(s)\tilde q(s)\tilde \psi(s)}{1-\tilde \psi(s)\tilde Q(s)}.
    \label{eq:Zs}
\end{align}
By taking the inverse Laplace transform of $\tilde Z(s)$ and then substituting it into Eq.~\eqref{eq:acf_simple}, one can obtain the analytical solution of the ACF for $x(t)$.

To demonstrate the above analytical results, we consider a simple case in which both reduced IETs and burst sizes are real numbers and exponentially distributed. By assuming that
\begin{align}
    \label{eq:expo_psi}
    \psi(\bar\tau)=\frac{1}{\langle\bar\tau\rangle}e^{-\bar\tau/\langle\bar\tau\rangle}
\end{align}
and
\begin{align}
    Q(b)=\frac{1}{\langle b\rangle}e^{-b/\langle b\rangle},
    \label{eq:expo_Q}
\end{align}
we derive an exact solution of the ACF as follows (see Appendix~\ref{append:expo}):
\begin{align}
    A(t_{\rm d})= e^{-t_{\rm d}/t_{\rm c}}\ \textrm{with}\ 
    t_{\rm c}\equiv 
    \frac{\langle\bar\tau\rangle \langle b\rangle}{\langle\bar\tau\rangle+\langle b\rangle}.
\end{align}

\section{Power-law case}\label{sec:powerlaw}

Let us return to our original question on temporal scaling behavior. We assume continuous versions of power-law distributions of reduced IETs and burst sizes as follows:
\begin{align}
    \label{eq:psi_cont_define}
    &\psi(\bar\tau)=C_\psi \bar\tau^{-\alpha}\ \textrm{for}\ 1\leq \bar\tau\leq \bar\tau_{\rm c},\\
    \label{eq:Q_cont_define}
    &Q(b)=C_Q b^{-\beta}\ \textrm{for}\ 1\leq b\leq b_{\rm c}.
\end{align}
Here $\alpha,\beta>1$ are power-law exponents, and $\bar\tau_{\rm c}$ and $b_{\rm c}$ are cutoffs. Also, $C_\psi \equiv\frac{\alpha-1}{1-\bar\tau_{\rm c}^{1-\alpha}}$ and $C_Q \equiv \frac{\beta-1}{1-b_{\rm c}^{1-\beta}}$ are normalization constants. Then we will derive the analytical result of the decay exponent $\gamma$ of the ACF as a function of $\alpha$ and $\beta$, i.e., $\gamma(\alpha,\beta)$.

We first prove a useful property that $\gamma$ is symmetric with respect to the exchange of $\alpha$ and $\beta$, namely,
\begin{align}
 \gamma(\alpha,\beta)=\gamma(\beta,\alpha).
 \label{eq:symmetry_gamma}
\end{align}
To prove this property, let us consider a complementary event sequence $y(t)$ to the original event sequence $x(t)$, which is defined as
\begin{align}
    y(t)\equiv 1-x(t).
    \label{eq:y_x}
\end{align}
The ACF defined for $y(t)$ using the formula in Eq.~\eqref{eq:acf_define} turns out to be the same as the ACF for $x(t)$:
\begin{align}
    \frac{ \langle y(t)y(t+t_{\rm d})\rangle_t- \langle y(t)\rangle^2_t}{ \langle y(t)^2\rangle_t- \langle y(t)\rangle^2_t}=
    \frac{ \langle x(t)x(t+t_{\rm d})\rangle_t- \langle x(t)\rangle^2_t}{ \langle x(t)^2\rangle_t- \langle x(t)\rangle^2_t}.
    \label{eq:acf_y}
\end{align}
That is, the decay exponent of the ACF for $x(t)$ must be the same as that for $y(t)$. By the definition of $y(t)$ in Eq.~\eqref{eq:y_x}, the periods of $x(t)=0$ correspond to those of $y(t)=1$ and vice versa. It means that reduced IETs and burst sizes in $x(t)$ respectively correspond to burst sizes and reduced IETs in $y(t)$, closing the proof.

Under Eq.~\eqref{eq:psi_cont_define}, the average of $\bar\tau$ is given by
\begin{align}
    \label{eq:taubar_avg}
    \langle\bar\tau\rangle \equiv \int_1^{\bar\tau_{\rm c}} \text{d}\bar\tau \bar\tau \psi(\bar\tau) =\begin{cases}
    \frac{\alpha-1}{2-\alpha}\frac{\bar\tau_{\rm c}^{2-\alpha}-1}{1-\bar\tau_{\rm c}^{1-\alpha}}
    & \textrm{for}\ \alpha\neq 2,\\
    \frac{\bar\tau_{\rm c}}{\bar\tau_{\rm c}-1}\ln \bar\tau_{\rm c}
    & \textrm{for}\ \alpha=2.
    \end{cases}
\end{align}
Similarly, under Eq.~\eqref{eq:Q_cont_define}, the average of $b$ reads
\begin{align}
    \label{eq:b_avg}
    \langle b\rangle \equiv \int_1^{b_{\rm c}} \text{d}b b Q(b) =\begin{cases}
    \frac{\beta-1}{2-\beta}\frac{b_{\rm c}^{2-\beta}-1}{1-b_{\rm c}^{1-\beta}}
    & \textrm{for}\ \beta\neq 2,\\
    \frac{b_{\rm c}}{b_{\rm c}-1}\ln b_{\rm c}
    & \textrm{for}\ \beta=2.
    \end{cases}
\end{align}
Note that the event rate $\lambda$ in Eq.~\eqref{eq:lambda} is determined by the above $\langle\bar\tau\rangle$ and $\langle b\rangle$.

We now divide the entire range of $\alpha,\beta>1$ into several cases to derive the analytical solution of the ACF in each case. Considering the symmetric nature of $\gamma$ in Eq.~\eqref{eq:symmetry_gamma}, the following cases are sufficient to get the complete picture of the result.

\subsection{Case with $\alpha=\beta=2$}

When $\alpha=\beta=2$, we get the Laplace transforms of $\psi(\bar\tau)$ in Eq.~\eqref{eq:psi_cont_define} and $Q(b)$ in Eq.~\eqref{eq:Q_cont_define} in the limit of $s\to 0$ as (see Appendix~\ref{append:Qs})
\begin{align}
    &\tilde \psi(s)\simeq 1+s\ln s,\\
    \label{eq:Qs_beta2}
    &\tilde Q(s)\simeq 1+s\ln s.
\end{align}
By substituting Eq.~\eqref{eq:Qs_beta2} in Eqs.~\eqref{eq:p0s}, \eqref{eq:Rs}, and \eqref{eq:qs}, we obtain
\begin{align}
    &\tilde p_0(s) = \frac{1}{s}\left(1+\frac{\ln s}{\langle b\rangle}\right),
    \label{eq:p0s_approx_b2}\\
    &\tilde R(s) = -\frac{\ln s}{\langle b\rangle},
    \label{eq:Rs_approx_b2}\\
    &\tilde q(s) = -\ln s.
    \label{eq:qs_approx_b2}
\end{align}
Using Eq.~\eqref{eq:Zs}, after some algebra, one obtains 
\begin{align}
    \tilde Z(s)\approx \frac{1}{s}+\frac{\ln s}{2\langle b\rangle s},
\end{align}
where $\approx$ represents ``approximately equal to'', leading to its inverse Laplace transform as
\begin{align}
    Z(t_{\rm d})\approx 1- \frac{\ln t_{\rm d}+\gamma_{_{\rm EM}}}{2\langle b\rangle}
\end{align}
with $\gamma_{_{\rm EM}}\approx 0.5772$ denoting the Euler-Mascheroni constant~\cite{Olver2024NIST}. Using Eq.~\eqref{eq:acf_simple} one obtains
\begin{align}
    A(t_{\rm d})\approx 1 - \frac{\langle\bar\tau\rangle + \langle b\rangle}{2\langle\bar\tau\rangle \langle b\rangle}(\ln t_{\rm d}+\gamma_{_{\rm EM}}),
    \label{eq:acf_a2b2}
\end{align}
enabling us to conclude that 
\begin{align}
    \gamma=0\ \textrm{for}\ \alpha=\beta=2.
\end{align}

\subsection{Case with $\alpha\neq 2$, $\beta=2$}

The Laplace transform of $\psi(\bar\tau)$ for $\alpha\neq 2$ is written as
\begin{align}
    \tilde\psi(s)=C_\psi s^{\alpha-1}\left[\Gamma(1-\alpha,s)-\Gamma(1-\alpha,s\bar\tau_{\rm c})\right],
\end{align}
where $\Gamma(\cdot,\cdot)$ is the upper incomplete Gamma function. For the intermediate range of $s$, i.e., $\frac{1}{\bar\tau_{\rm c}}\ll s\ll 1$, we obtain $C_\psi\approx \alpha-1$ and $\Gamma(1-\alpha,s\bar\tau_{\rm c})\approx 0$, resulting in
\begin{align}
    \tilde \psi(s)\approx (\alpha-1)s^{\alpha-1}\Gamma(1-\alpha,s).
    \label{eq:psis_asym}
\end{align}
We expand Eq.~\eqref{eq:psis_asym} in the limit of $s \to 0$ to obtain
\begin{align}
    \tilde \psi(s) = 1+a_{\rm s}s^{\alpha-1}+a_1s
    +\mathcal{O}(s^2),
    \label{eq:psis_approx}
\end{align}
where $a_{\rm s}\equiv (\alpha-1)\Gamma(1-\alpha)$ and $a_1\equiv \frac{\alpha-1}{2-\alpha}$. Here $\Gamma(\cdot)$ is the Gamma function. As for $\tilde Q(s)$ and other functions derived from $\tilde Q(s)$, we keep using Eqs.~\eqref{eq:Qs_beta2}--\eqref{eq:qs_approx_b2}. For $1<\alpha<2$, after some algebra, one obtains 
\begin{align}
    \tilde Z(s)\approx \frac{1}{s} +\frac{\ln s}{\langle b\rangle s},
\end{align}
implying
\begin{align}
    Z(t_{\rm d})\approx 1- \frac{\ln t_{\rm d}+\gamma_{_{\rm EM}}}{\langle b\rangle}.
\end{align}
Using Eq.~\eqref{eq:acf_simple} we obtain
\begin{align}
    A(t_{\rm d})\approx 1 - \frac{\langle\bar\tau\rangle + \langle b\rangle}{\langle\bar\tau\rangle \langle b\rangle}(\ln t_{\rm d}+\gamma_{_{\rm EM}}),
    \label{eq:acf_anot2b2}
\end{align}
hence
\begin{align}
    \gamma=0\ \textrm{for}\ 1<\alpha< 2,\ \beta=2.
\end{align}
For $\alpha>2$, we obtain
\begin{align}
    \tilde Z(s)\approx \frac{1}{s}-\frac{a_1}{\langle b\rangle}\ln s.
    \label{eq:no_solution}
\end{align}
Although the inverse Laplace transform of $\ln s$ does not exist, we can still conclude that 
\begin{align}
    \gamma=0\ \textrm{for}\ \alpha> 2,\ \beta=2.
\end{align}
Thanks to the symmetric property of $\gamma$ in Eq.~\eqref{eq:symmetry_gamma}, one concludes that
\begin{align}
    \gamma=0\ \textrm{for}\ \alpha=2\ \textrm{or}\ \beta=2.
\end{align}

\subsection{Case with $1<\alpha,\beta<2$}

We first study the case with $\alpha\leq \beta$ and then the solution in the case of $\alpha>\beta$ will be obtained via Eq.~\eqref{eq:symmetry_gamma}. Similarly to the case of $\tilde \psi(s)$ in Eqs.~\eqref{eq:psis_asym} and~\eqref{eq:psis_approx}, we get the expanded $\tilde Q(s)$ for the intermediate range of $s$, i.e., $\frac{1}{b_{\rm c}}\ll s\ll 1$, as follows:
\begin{align}
    \tilde Q(s) = 1+c_{\rm s}s^{\beta-1}+c_1s +c_2s^2+c_3s^3+\mathcal{O}(s^4),
    \label{eq:Qs_approx}
\end{align}
where $c_{\rm s}\equiv (\beta-1)\Gamma(1-\beta)$, $c_1\equiv \frac{\beta-1}{2-\beta}$, $c_2\equiv -\frac{\beta-1}{2(3-\beta)}$, and $c_3\equiv \frac{\beta-1}{6(4-\beta)}$. Again using Eqs.~\eqref{eq:p0s},~\eqref{eq:Rs}, and~\eqref{eq:qs}, we obtain
\begin{align}
    &\tilde p_0(s) = \left(1+\frac{c_1}{\langle b\rangle}\right)\frac{1}{s} + \frac{c_{\rm s}}{\langle b\rangle}s^{\beta-3} +\mathcal{O}(1),
    \label{eq:p0s_finite_approx}\\
    &\tilde R(s) = -\frac{c_{\rm s}}{\langle b\rangle} s^{\beta-2}+\mathcal{O}(1),
    \label{eq:Rs_finite_approx}\\
    &\tilde q(s) = - c_{\rm s}s^{\beta-2} + \mathcal{O}(1).
    \label{eq:qs_finite_approx}
\end{align}
For the case with $\alpha<\beta$, after some algebra, we obtain $\tilde Z(s)$ in Eq.~\eqref{eq:Zs} up to the leading terms as follows:
\begin{align}
    \tilde Z(s)\approx 
    \left(1+\frac{c_1}{\langle b\rangle}\right)\frac{1}{s} + \frac{c_{\rm s}}{\langle b\rangle}s^{\beta-3}
    -\frac{c_{\rm s}^2}{a_{\rm s}\langle b\rangle}s^{2\beta-\alpha-3}.
\label{eq:Zs_expanded_finite}
\end{align}
Obviously, the last term on the right-hand side of Eq.~\eqref{eq:Zs_expanded_finite} is dominated by the second term. We find that for the intermediate range of $s$, specifically, $\frac{1}{b_{\rm c}}\ll s\ll 1$,
the second term is dominated by the first term because
\begin{align}
    \frac{s^{\beta-3}}{\langle b\rangle}\propto (sb_{\rm c})^{\beta-2}\frac{1}{s}\ll \frac{1}{s}.
\end{align}
Finally, for the first term in Eq.~\eqref{eq:Zs_expanded_finite}, since $c_1\ll \langle b\rangle$ for $b_{\rm c}\gg 1$, we obtain up to the second leading term
\begin{align}
    \tilde Z(s)\approx \frac{1}{s} + \frac{c_{\rm s}}{\langle b\rangle}s^{\beta-3}, 
\end{align}
which leads to
\begin{align}
Z(t_{\rm d})\approx 1 +\frac{c_{\rm s}}{\langle b\rangle}\frac{t_{\rm d}^{2-\beta}}{\Gamma(3-\beta)}.
\label{eq:Z-1-with-cutoff}
\end{align}
Note that the coefficient of the term $t_{\rm d}^{2-\beta}$ is negative for the range of $1<\beta<2$. By substituting Eq.~\eqref{eq:Z-1-with-cutoff} in Eq.~\eqref{eq:acf_simple}, we obtain
\begin{align}
    A(t_{\rm d})\approx 1 +
    \frac{\langle\bar\tau\rangle+\langle b\rangle}{\langle \bar\tau\rangle\langle b\rangle}\frac{c_{\rm s}t_{\rm d}^{2-\beta}}{\Gamma(3-\beta)}.
    \label{eq:acf_a12b12}
\end{align}
In the case with $\alpha=\beta$, we similarly obtain the ACF as follows:
\begin{align}
A(t_{\rm d})\approx 1 +
    \frac{\langle\bar\tau\rangle+\langle b\rangle}{2\langle \bar\tau\rangle\langle b\rangle}\frac{c_{\rm s}t_{\rm d}^{2-\beta}}{\Gamma(3-\beta)}.
    \label{eq:acf_ab12}
\end{align}
Therefore, we conclude that
\begin{align}
    \gamma=0\ \textrm{for}\ 1<\alpha\leq \beta<2.
    \label{eq:alpha12beta12_part}
\end{align}
Owing to the symmetric nature of $\gamma(\alpha,\beta)$ in Eq.~\eqref{eq:symmetry_gamma}, we further conclude that
\begin{align}
    \gamma=0\ \textrm{for}\ 1<\alpha,\beta<2.
    \label{eq:alpha12beta12}
\end{align}

\begin{figure*}[!t]
\includegraphics[width=\textwidth]{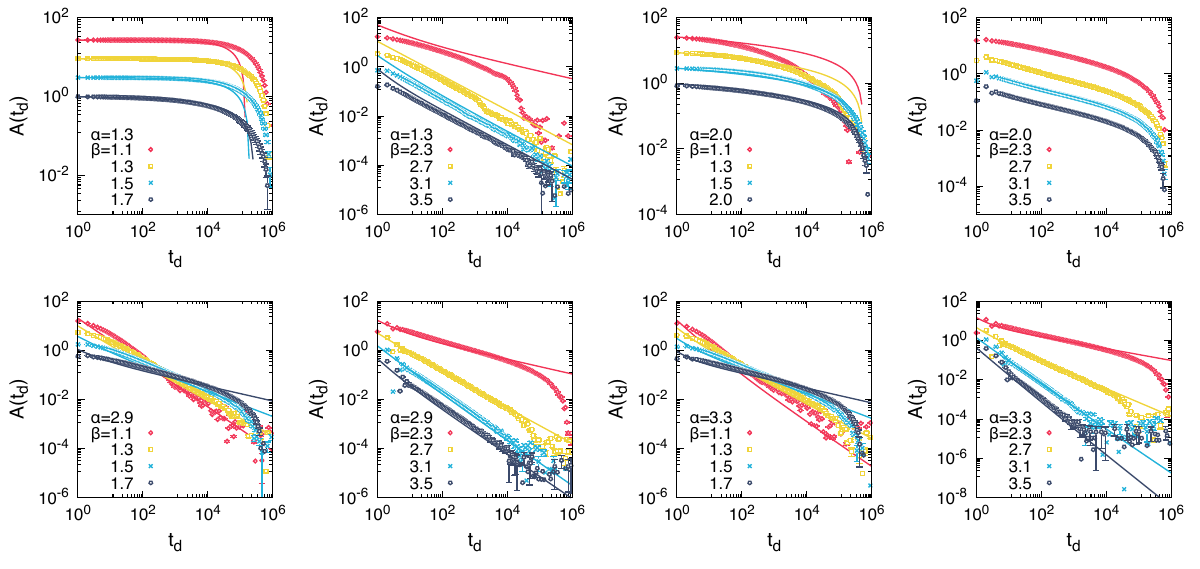}
\caption{Numerical autocorrelation functions $\tilde A(t_{\rm d})$ defined as Eq.~\eqref{eq:acf_numeric} (symbols) that are calculated for the time series $x(t)$ generated using the model with various combinations of $\alpha$ and $\beta$. Each ACF curve was obtained from $100$ event sequences generated with $T=5\times 10^7$ and $\bar\tau_{\rm c}=b_{\rm c}=10^6$. Error bars denote standard errors. These simulation results are compared with corresponding analytical solutions of $A(t_{\rm d})$ (solid lines), which are respectively Eqs.~\eqref{eq:acf_a2b2}, \eqref{eq:acf_anot2b2}, \eqref{eq:acf_a12b12}, \eqref{eq:acf_ab12}, \eqref{eq:acf_ab>2}, and~\eqref{eq:acf_other}, and some of these equations with $\alpha$ and $\beta$ being exchanged ($\bar\tau_{\rm c}$ and $b_{\rm c}$ being exchanged too whenever necessary). Note that there are no analytical solutions for the case with $\alpha=2$ and $\beta>2$ [Eq.~\eqref{eq:no_solution}]. In all panels, some curves are vertically shifted for better visualization.}
\label{fig:acf}
\end{figure*}

\subsection{Case with $\alpha,\beta>2$}

For the case with $\alpha,\beta>2$, we use the expanded $\tilde\psi(s)$ in Eq.~\eqref{eq:psis_approx} and the expanded $\tilde Q(s)$ in Eq.~\eqref{eq:Qs_approx} in the limit of $s\to 0$. Since $a_1=-\langle \bar\tau\rangle$ for $\alpha>2$ and $\bar\tau_{\rm c}\gg 1$ [Eq.~\eqref{eq:taubar_avg}], we replace $a_1$ by $-\langle \bar\tau\rangle$. Similarly, we replace $c_1$ by $-\langle b\rangle$. Using Eqs.~\eqref{eq:p0s},~\eqref{eq:Rs}, and~\eqref{eq:qs}, we obtain
\begin{align}
    &\tilde p_0(s) = \frac{c_2}{\langle b\rangle} +\frac{c_{\rm s}}{\langle b\rangle}s^{\beta-3} + \frac{c_3}{\langle b\rangle}s +\mathcal{O}(s^2),
    \label{eq:p0s_approx}\\
    &\tilde R(s) = 1-\frac{c_{\rm s}}{\langle b\rangle} s^{\beta-2}-\frac{c_2}{\langle b\rangle}s +\mathcal{O}(s^2),
    \label{eq:Rs_approx}\\
    &\tilde q(s) = \langle b\rangle - c_{\rm s}s^{\beta-2} - c_2 s + \mathcal{O}(s^2).
    \label{eq:qs_approx}
\end{align}
After some algebra, we derive $\tilde Z(s)$ in Eq.~\eqref{eq:Zs} up to the leading terms as
\begin{align}
    \tilde Z(s)\approx 
    &\frac{\langle b\rangle}{\langle\bar\tau\rangle+\langle b\rangle}\frac{1}{s}
    +\frac{a_{\rm s}\langle b\rangle}{(\langle\bar\tau\rangle+\langle b \rangle)^2}s^{\alpha-3}\notag\\
    &+\frac{c_{\rm s}\langle\bar\tau\rangle^2}{\langle b\rangle(\langle\bar\tau\rangle+\langle b\rangle)^2}s^{\beta-3},
\end{align}
leading to its inverse Laplace transform as
\begin{align}
    Z(t_{\rm d})\approx 
    &\frac{\langle b\rangle}{\langle\bar\tau\rangle+\langle b\rangle}
    +\frac{a_{\rm s}\langle b\rangle}{(\langle\bar\tau\rangle+\langle b\rangle)^2}\frac{t_{\rm d}^{-(\alpha-2)}}{\Gamma(3-\alpha)}\notag\\
    &+\frac{c_{\rm s}\langle\bar\tau\rangle^2}{\langle b\rangle(\langle\bar\tau\rangle+\langle b\rangle)^2}\frac{t_{\rm d}^{-(\beta-2)}}{\Gamma(3-\beta)}.
    \label{eq:Ztd_alpha2beta2}
\end{align}
Since the constant term on the right hand side of Eq.~\eqref{eq:Ztd_alpha2beta2} is cancelled with $\lambda$ in Eq.~\eqref{eq:lambda}, one obtains from Eq.~\eqref{eq:acf_simple}
\begin{align}
    A(t_{\rm d})\approx 
    \frac{a_{\rm s}\langle b\rangle}{\langle\bar\tau\rangle
    (\langle\bar\tau\rangle+\langle b\rangle)}\frac{t_{\rm d}^{-(\alpha-2)}}{\Gamma(3-\alpha)}
    +\frac{c_{\rm s}\langle\bar\tau\rangle}{\langle b\rangle(\langle\bar\tau\rangle+\langle b\rangle)}\frac{t_{\rm d}^{-(\beta-2)}}{\Gamma(3-\beta)},
    \label{eq:acf_ab>2}
\end{align}
enabling us to find the scaling relation: 
\begin{align}
    \gamma=\min\{\alpha-2,\beta-2\}\ \textrm{for}\ \alpha,\beta>2.
    \label{eq:alpha2beta2}
\end{align}
Note that this $\gamma$ is symmetric with respect to the exchange of $\alpha$ and $\beta$.

\subsection{Case with $1<\alpha<2$, $\beta>2$}

When $1<\alpha<2$ and $\beta>2$, by combining the expanded $\tilde\psi(s)$ given by Eq.~\eqref{eq:psis_approx}, the expanded $\tilde Q(s)$ given by Eq.~\eqref{eq:Qs_approx}, and related functions given by Eqs.~\eqref{eq:p0s_approx}--\eqref{eq:qs_approx}, we obtain up to the leading terms
\begin{align}
    \tilde Z(s)\approx 
    % \frac{c_2}{\langle b\rangle}+ 
    \frac{c_{\rm s}}{\langle b\rangle}s^{\beta-3}-\frac{\langle b\rangle}{a_{\rm s}}s^{1-\alpha}.
    \label{eq:Zs_alpha12}
\end{align}
The inverse Laplace transform of Eq.~\eqref{eq:Zs_alpha12} results in
\begin{align}
    Z(t_{\rm d})\approx \frac{c_{\rm s}}{\langle b\rangle}\frac{t_{\rm d}^{-(\beta-2)}}{\Gamma(3-\beta)}-\frac{\langle b\rangle}{a_{\rm s}}\frac{t_{\rm d}^{-(2-\alpha)}}{\Gamma(\alpha-1)}.
    \label{eq:acf_other}
\end{align}
Since $\langle \bar\tau\rangle$ diverges for $\alpha<2$ and $\bar\tau_{\rm c}\gg 1$, one obtains the vanishing event rate, i.e., $\lambda=0$ [Eq.~\eqref{eq:lambda}]. Thus, $A(t_{\rm d})=Z(t_{\rm d})$ from Eq.~\eqref{eq:acf_simple}, and we get the scaling relation:
\begin{align}
    \gamma=\min\{2-\alpha,\beta-2\}\ \textrm{for}\ 1<\alpha<2,\ \beta>2.
    \label{eq:alpha12beta2}
\end{align}
Again thanks to the symmetric nature of $\gamma$, we conclude that
\begin{align}
    \gamma=\min\{\alpha-2,2-\beta\}\ \textrm{for}\ \alpha>2,\ 1<\beta<2.
    \label{eq:alpha2beta12}
\end{align}

In sum, we have derived the power-law exponent $\gamma$ of the ACF decay as a function of the IET power-law exponent $\alpha$ and burst-size power-law exponent $\beta$ for the entire range of $\alpha,\beta>1$. We summarize the results in Eq.~\eqref{eq:gamma_summary} and depict them in Fig.~\ref{fig:main_result}.

\subsection{Numerical simulation}

To test the validity of our analytical solution given by Eq.~\eqref{eq:gamma_summary}, we generate the event sequence $\{ x(1), \ldots, x(T) \}$ using the following distributions of reduced IETs and burst sizes:
\begin{align}
    \label{eq:Ptau_numeric}
    &\psi(\bar\tau)=C_\psi \bar\tau^{-\alpha}\ \textrm{for}\ \bar\tau=1,2,\ldots, \bar\tau_{\rm c},\\
    \label{eq:Qb_numeric}
    &Q(b)=C_Q b^{-\beta}\ \textrm{for}\ b=1,2,\ldots, b_{\rm c},
\end{align}
where $C_\psi=(\sum_{\bar\tau=1}^{\bar\tau_{\rm c}} \bar\tau^{-\alpha})^{-1}$ and $C_Q=(\sum_{b=1}^{b_{\rm c}} b^{-\beta})^{-1}$. Precisely, we randomly draw a burst size $b_1$ from $Q(b)$ in Eq.~\eqref{eq:Qb_numeric} to set $x(t)=1$ for $t=1,\ldots,b_1$. Then a reduced IET $\bar\tau_1$ is randomly drawn from $\psi(\bar\tau)$ in Eq.~\eqref{eq:Ptau_numeric} to set $x(t)=0$ for $t=b_1+1,\ldots,b_1+\bar\tau_1$. We draw another burst size $b_2$ from $Q(b)$ and another reduced IET $\bar\tau_2$ from $\psi(\bar\tau)$, respectively, to set $x(t)=1$ for $t=b_1+\bar\tau_1+1,\ldots,b_1+\bar\tau_1+b_2$ and $x(t)=0$ for $t=b_1+\bar\tau_1+b_2+1,\ldots,b_1+\bar\tau_1+b_2+\bar\tau_2$. We repeat this procedure until $t$ reaches $T$.

Using the generated time series $\{ x(1), \ldots, x(T) \}$, we numerically calculate the ACF by
\begin{align}
    \tilde A(t_{\rm d})\equiv \frac{\frac{1}{T-t_{\rm d}} \sum_{t=1}^{T-t_{\rm d}} x(t)x(t+t_{\rm d})-\lambda_1\lambda_2}{\sigma_1\sigma_2},
    \label{eq:acf_numeric}
\end{align}
where $\lambda_1$ and $\sigma_1$ are respectively the average and standard deviation of $\{x(1), \ldots, x(T-t_{\rm d})\}$, and $\lambda_2$ and $\sigma_2$ are respectively the average and standard deviation of $\{x(t_{\rm d}+1),\ldots, x(T)\}$. 

For the simulations, we use $T=5\times 10^7$ and $\bar\tau_{\rm c}=b_{\rm c}=10^6$ to generate $100$ different event sequences $x(t)$ for each combination of $\alpha$ and $\beta$. Then their autocorrelation functions are calculated using Eq.~\eqref{eq:acf_numeric}. As shown in Fig.~\ref{fig:acf}, simulation results in terms of $\tilde A(t_{\rm d})$ for several combinations of parameters of $\alpha$ and $\beta$ are in good agreement with corresponding analytical solutions of $A(t_{\rm d})$ in most cases. In some cases we observe systematic deviations of analytical solutions from the simulation results, which may be due to ignorance of higher-order terms when deriving analytical solutions.

\section{Conclusion}\label{sec:conclusion}

To study the combined effects of the interevent time (IET) distribution $P(\tau)$ and the burst size distribution $Q(b)$ on the autocorrelation function (ACF) $A(t_{\rm d})$, we have devised a model for generating time series using $P(\tau)$ and $Q(b)$ as inputs. Our model is simple but takes correlations between an arbitrary number of consecutive IETs into account in terms of bursty trains~\cite{Karsai2012Universal}. We are primarily interested in temporal scaling behaviors observed in $A(t_{\rm d})\propto t_{\rm d}^{-\gamma}$ when $P(\tau)\propto \tau^{-\alpha}$ (except at $\tau=1$) and $Q(b)\propto b^{-\beta}$ are assumed. We have derived the analytical solutions of $A(t_{\rm d})$ for arbitrary values of IET power-law exponent $\alpha>1$ and burst-size power-law exponent $\beta>1$ to obtain the ACF decay power-law exponent $\gamma$ as a function of $\alpha$ and $\beta$ [Eq.~\eqref{eq:gamma_summary}; see also Fig.~\ref{fig:main_result}].

We remark that our model has assumed that IETs with $\tau\geq 2$ and burst sizes in the time series are independent of each other. However, there are observations indicating the presence of correlations between consecutive burst sizes and even higher-order temporal structure~\cite{Jo2020Bursttree}. Thus, it would be interesting to see whether such higher-order structure affects the decaying behavior of the ACF. 

So far, we have focused on the analysis of the single time series observed for a single phenomenon or for a system as a whole. However, there are other complex systems in which each element of the system has its own bursty activity pattern or a pair of elements have their own bursty interaction pattern, such as calling patterns between mobile phone users~\cite{Jo2012Circadian, Saramaki2015Seconds}. In recent years, such systems have been studied in the framework of temporal networks~\cite{Holme2012Temporal, Holme2023Temporal, Masuda2021Guide}, where interaction events are heterogeneously distributed among elements as well as over the time axis. Modeling temporal networks is important to understand collective dynamics, such as spreading or diffusion~\cite{Pastor-Satorras2015Epidemic}, taking place in those networks. Some recent efforts for modeling temporal networks are mostly concerned with heavy-tailed IET distributions for each element or a pair of elements~\cite{Hiraoka2020Modeling, Sheng2023Constructing}. Our model can be extended to generate more realistic temporal networks in which activity patterns of elements or interaction patterns between elements are characterized by bursty time series with higher-order temporal structure beyond the IET distribution.

\begin{acknowledgments}
H.-H.J. and T.B. acknowledge financial support by the National Research Foundation of Korea (NRF) grant funded by the Korea government (MSIT) (No. 2022R1A2C1007358).
N.M. acknowledges financial support by the Japan Science and Technology Agency (JST) Moonshot R\&D (under Grant No. JPMJMS2021), the National Science Foundation (under Grant Nos. 2052720 and 2204936), and JSPS KAKENHI (under grant Nos. JP 21H04595 and
23H03414).
\end{acknowledgments}

\appendix

\section{Derivation of the normalization constant in Eq.~\eqref{eq:R_define}}\label{append:CR}

Let us write $R(r)$ as follows:
\begin{align}
    R(r)= C_R\sum_{b=r+1}^\infty Q(b).
    \label{eq:R_define_appendix}
\end{align}
Then we derive the normalization constant $C_R$ from the normalization condition for $R(r)$:
\begin{align}
    1=\sum_{r=0}^\infty R(r)= C_R \sum_{r=0}^\infty \sum_{b=r+1}^\infty Q(b).
\end{align}
Two summations on the right hand side can be exchanged as
\begin{align}
    \sum_{r=0}^\infty \sum_{b=r+1}^\infty = \sum_{b=1}^\infty \sum_{r=0}^{b-1}\, ,
\end{align}
leading to
\begin{align}
    1=C_R \sum_{b=1}^\infty \sum_{r=0}^{b-1} Q(b) = C_R \sum_{b=1}^\infty bQ(b)=C_R \langle b\rangle.
\end{align}
Finally one obtains
\begin{align}
    C_R=\frac{1}{\langle b\rangle}.
\end{align}

\section{Analysis for the case with exponential distributions of reduced IETs and burst sizes}\label{append:expo}

By substituting $Q(b)$ in Eq.~\eqref{eq:expo_Q} in Eqs.~\eqref{eq:R_cont_define}, \eqref{eq:Rf_cont_define}, and \eqref{eq:p_0_cont}, we obtain
\begin{align}
\label{eq:expo_Rr}
    &R(r)=\frac{1}{\langle b\rangle}e^{-r/\langle b\rangle},\\
    &q(r')=e^{-r'/\langle b\rangle},\\
    &p_0(t_{\rm d})=e^{-t_{\rm d}/\langle b\rangle}.
    \label{eq:expo_p0}
\end{align}
We take the Laplace transform of Eqs.~\eqref{eq:expo_psi},~\eqref{eq:expo_Q} and \eqref{eq:expo_Rr}--\eqref{eq:expo_p0} to obtain
\begin{align}
    &\tilde\psi(s)=\frac{1}{\langle\bar\tau\rangle s+1},\\
    &\tilde Q(s)=\tilde R(s)=\frac{1}{\langle b\rangle s+1},\\
    &\tilde q(s)=\tilde p_0(s)=\frac{\langle b\rangle }{\langle b\rangle s+1}.
\end{align}
Using Eq.~\eqref{eq:Zs}, one obtains
\begin{align}
    \tilde Z(s) = \frac{\langle b\rangle }{\langle b\rangle s+1}\left[1 + \frac{1}{s(\langle b\rangle \langle\bar\tau\rangle s+\langle\bar\tau\rangle +\langle b\rangle)}\right].
\label{eq:tildeZ(s)-exponential}
\end{align}
The inverse Laplace transform of Eq.~\eqref{eq:tildeZ(s)-exponential} is given by
\begin{align}
    Z(t_{\rm d})= \frac{\langle b\rangle}{\langle\bar\tau\rangle+\langle b\rangle} + \frac{\langle\bar\tau\rangle e^{-t_{\rm d}(\langle\bar\tau\rangle+\langle b\rangle)/(\langle\bar\tau\rangle \langle b\rangle)}}{\langle\bar\tau\rangle+\langle b\rangle},
\end{align}
which leads to
\begin{align}
    A(t_{\rm d})= e^{-t_{\rm d}/t_{\rm c}},
\end{align}
where
\begin{align}
    t_{\rm c}\equiv 
    \frac{\langle\bar\tau\rangle \langle b\rangle}{\langle\bar\tau\rangle+\langle b\rangle}.
\end{align}

\section{Derivation of the Laplace transforms of $\psi(\bar\tau)$ with $\alpha=2$ and $Q(b)$ with $\beta=2$}\label{append:Qs}

We take the Laplace transform of $\psi(\bar\tau)$ with $\alpha=2$ in Eq.~\eqref{eq:psi_cont_define}:
\begin{align}
    \tilde\psi(s)=C_\psi \int_1^{\bar\tau_{\rm c}} \text{d}\bar\tau e^{-s\bar\tau}\bar\tau^{-2},
\end{align}
where $C_\psi=\frac{\bar\tau_{\rm c}}{\bar\tau_{\rm c}-1}$. For $\bar\tau_{\rm c}\gg 1$, one obtains $C_\psi\approx 1$. Then,
\begin{align}
    1-\tilde\psi(s)=\int_1^{\bar\tau_{\rm c}} \text{d}\bar\tau (1-e^{-s\bar\tau})\bar\tau^{-2}.
\end{align}
The change of the integrated variable from $\bar{\tau}$ to $v \equiv s\bar\tau$ yields
\begin{align}
    1-\tilde\psi(s)=s \int_s^{\infty} \text{d}v (1-e^{-v})v^{-2}.
\end{align}
We have assumed that $s\bar\tau_{\rm c}\gg 1$. We take the limit of $s\to 0$ after dividing $1-\tilde\psi(s)$ by $s\ln s$:
\begin{align}
    \lim_{s\to 0}\frac{1-\tilde\psi(s)}{s\ln s}&=\lim_{s\to 0}\frac{\int_s^{\infty} \text{d}v (1-e^{-v})v^{-2}}{\ln s}\\
    &=\lim_{s\to 0}\frac{-(1-e^{-s})s^{-2}}{1/s}\\
    &=\lim_{s\to 0}\frac{-(1-e^{-s})}{s}\\
    &=\lim_{s\to 0}-e^{-s}=-1,
\end{align}
where we have used L'H\^opital's rule for the derivation~\cite{Lowen2005Fractalbased}. Therefore, in the limit of $s\to 0$, we obtain
\begin{align}
    \tilde\psi(s)\approx 1+s\ln s.
\end{align}
Similarly, we obtain $\tilde Q(s)\approx 1 + s\ln s$.

%\bibliography{h2jo-papers}% Produces the bibliography via BibTeX.

\end{document}